\documentstyle[twocolumn,aps,epsf]{revtex}

\renewcommand{\vec}[1]{\mbox{\boldmath$#1$\unboldmath}}
\begin{document}

\title{Momentum Distributions in Halo Nuclei}
\author{J\"urgen Wurzer\thanks{e-mail:
{\tt jwurzer@theorie3.physik.uni-erlangen.de}} and Hartmut M.\
Hofmann}
\address{Institut f\"ur Theoretische Physik III, Staudtstra{\ss}e 7,
 Universit\"at
Erlangen-N\"urnberg, D 91058 Erlangen, Germany}
\maketitle

\widetext
\begin{abstract}
The halo nuclei  $^6$He and $^8$He are described in a consistent way in
a microscopic multiconfiguration model, the refined resonating group
method. The ground state properties have been calculated, and momentum
distributions of fragments and neutrons have been determined in a simple
reaction scenario, taking into account final-state interactions.
The correlation of neutrons and fragments are investigated.

\vspace{12pt}
{\parindent0.0pt PACS-numbers: 21.60.-n, 21.60.Gx, 27.20.+n}
\end{abstract}
\narrowtext

\section{Introduction}

\begin{samepage}
Nuclear halos -- a meanwhile well established phenomena of light neutron rich
nuclei -- are considered to be roughly understood; and at least the ground
state structure of some halo nuclei like $^{11}$Be and especially
$^6$He are even thought to be almost completely enlightened.

Also the measured momentum distributions of halo nuclei fragments can be
described very well within various theoretical models [1-7].
But in all
these models some free parameters have been adjusted to reproduce the
properties of one nucleus. Contrary to this our aim is a consistent 
description of a large number of light nuclei, including halo nuclei,
within one model whose parameters are fixed once and for all.

First results of this project have been presented in
\cite{Wu97} for a consistent description 
of the nuclei
$^4$He -- $^8$He. It turned out that the results for increasing
mass number become more and more sensitive to the $P$-wave part of the
employed effective nucleon-nucleon interaction. This part 
of the potential leads to an overbound $^8$He indicating a lack of repulsion 
and hence yielding a too small radius.

In order to cure this problem we removed the odd partial wave part of the
effective potential and added a realistic $NN$ interaction with a more 
repulsive core.

Here we show that with this modified interaction in addition to the lighter 
isotopes also the heavier isotopes can be described very well in a consistent
way, without adjusting any parameter of the interaction to the individual
systems.

Besides the ground state properties we compare momentum distributions of 
fragments and neutrons from $^6$He and $^8$He breakup -- calculated in sudden 
approximation and taking into account final-state interactions --
with experimental data and calculations in other models. 
We also discuss the correlation of the halo neutrons.
\end{samepage}
 
\mbox{}
\vspace*{2.254cm}
\section{Spatial Structure} 

All calculations have been performed in a microscopic multiconfiguration
cluster model -- the refined resonating group method (RRGM).
Details of the model are given in \cite{Wu97}.

The $^6$He wave function is a superposition of the (nonorthogonal)
clusterings
$\alpha(nn)$, $(\alpha n)n$ and $tt$ corresponding to the $^4$He+2$n$,
$^5$He+$n$ and $^3$H+$^3$H thresholds of the $^6$He system.
On the cluster relative coordinates
orbital angular momenta $l=0,1$ have been allowed for
the $^6$He ground state. Between the nuclei
inside the $\alpha$ and triton clusters only $l=0$ are considered.
Four Gaussian functions were employed for the description of each cluster
relative wave function and one Gaussian function for the $\alpha$ and the
triton cluster each. All Gaussian width parameters have been determined 
by variation.

\vbox{
\begin{table}
\caption{\label{he6gsp}Calculated binding energy $E_{\rm th}$ with respect 
	 to the $^4$He+2$n$ threshold, matter $r_{\rm m}$ and charge 
         $r_{\rm ch}$ r.m.s.\ point nucleon radii for $^6$He compared with 
         experimental data.} 
\begin{tabular}{lr@{.}lr@{.}lr@{.}l}
 & \multicolumn{2}{c}{$E_{\rm th}$ [MeV]} & 
    \multicolumn{2}{c}{$r_{\rm m}$ [fm]}  & 
    \multicolumn{2}{c}{$r_{\rm ch}$ [fm]} \\
\hline
 Experiment & 0 & 975$^{\cite{Aj88}}$ & 2 & 33$\pm$0.04$^{\cite{Ta92}}$ &
 1 & 72$\pm$0.04$^{\cite{Ta92}}$ \\
 & \multicolumn{2}{c}{} & 2 & 30$\pm$0.07$^{\cite{Al97}}$ & 
   \multicolumn{2}{c}{} \\
 RRGM & 0 & 85 & 2 & 47 & 1 & 84 \\
\end{tabular}
\end{table}
}

In Table \ref{he6gsp} the calculated binding energy with respect to the first
threshold and the matter and charge radii are compared with experimental
data. The binding energy is about 0.1 MeV to small and hence the radii 
are a bit too large.

Table \ref{he6omit} shows the remaining binding energy for the case that
one omits each one of the
configurations in order to determine the importance of each configuration.
Without the $(\alpha n)n$ clustering with $P$-waves on both cluster
relative coordinates (the $^5$He g.s.\ + $n$ configuration) the binding is 
lost. Omitting the triton-triton
clustering decreases the binding considerable, which is surprising because
of the high threshold but in 
agreement with the calculations of Cs\'{o}t\'{o} \cite{Cs93}. Omitting any 
one of the other
configurations effects the binding energy hardly at all.

\vbox{
\begin{table}
\caption{\label{he6omit}Remaining binding energy for $^6$He omitting 
	 each one of the configurations.}
\begin{tabular}{lr@{=}lc}
 \multicolumn{3}{c}{omitted configuration} & $E_{\rm th}$ [MeV] \\
\hline
 none & \multicolumn{2}{c}{} & 0.85 \\
 $(\alpha n)n$ & $l_{1,2}$ & $0$ & 0.81 \\
 $(\alpha n)n$ & $l_{1,2}$ & $1$  & unbound \\
 tt            & $l$ & $0$ & 0.22 \\
 $\alpha(nn)$  & $l_{1,2}$ & $0$ & 0.81 \\
 $\alpha(nn)$  & $l_{1,2}$ & $1$ & 0.67 \\
\end{tabular}
\end{table}
}

The $P$-state probability is 14.1 \% and thus in agreement with a
hyperspherical harmonics calculation \cite{Zh91}.

The $^8$He wave function is a superposition of the (nonorthogonal) clusterings
$(\alpha(nn))(nn)$ with $l=0$ on all coordinates,
$(((\alpha n)n)n)n$ with $l=1$ on all coordinates,
$((\alpha n)n)(nn)$ with $l=1$ to the $n$ clusters and $l=0$ to the $nn$ 
cluster,
$((\alpha(nn))n)n$ with $l=1$ to the $n$ clusters and $l=0$ to the $nn$ cluster,
and $((tt)n)n$ with $l=0$ on all coordinates and in addition also $l=0$ between
the $t$ clusters and $l=1$ to the $n$ clusters.

The calculated binding energy with respect to the 
$^4$He+4$n$ threshold and the matter and charge radii are in good 
agreement with experimental data (cf.\ Table \ref{he8gsp}).

\vbox{
\begin{table}
\caption{\label{he8gsp}Calculated binding energy $E_{\rm th}$ with 
		respect to the $^4$He+4$n$ threshold, matter $r_{\rm m}$ 
		and charge $r_{\rm ch}$ r.m.s.\ point nucleon radii for 
		$^8$He compared with experimental data.} 
\begin{tabular}{lr@{.}lr@{.}lr@{.}l}
 & \multicolumn{2}{c}{$E_{\rm th}$ [MeV]} & 
   \multicolumn{2}{c}{$r_{\rm m}$ [fm]}   & 
   \multicolumn{2}{c}{$r_{\rm ch}$ [fm]}  \\
\hline
 Experiment & 3 & 112$^{\cite{Aj88}}$ & 2 & 49$\pm$0.04$^{\cite{Ta92}}$ &
 1 & 76$\pm$0.03$^{\cite{Ta92}}$ \\
 & \multicolumn{2}{c}{} & 2 & 45$\pm$0.07$^{\cite{Al97}}$ & 
 \multicolumn{2}{c}{} \\
 RRGM & 2 & 99 & 2 & 41 & 1 & 71  \\
\end{tabular}
\end{table}
}

The most important configuration for the binding energy is 
$(((\alpha n)n)n)n$ 
(cf.\ Table \ref{he8omit}). The contribution of the $((tt)n)n$ channel
is substantial.
The $P$-state probability is 49.8 \%. We omitted the $D$-states totally
since it turned out that they contribute only a few keV to the
binding energy.

Both wave functions are determined with the same $NN$ interaction by a
variational principle.
For the even partial waves the effective potential of St\"owe and Zahn
\cite{Me86},

\vbox{
\begin{table}
\caption{Remaining binding energy with respect to the $^4$He+4$n$ threshold 
	 for $^8$He omitting each one of the configurations.}
\label{he8omit}
\begin{tabular}{lc}
omitted configuration&   $E_{\rm th}$ [MeV] \\
\hline
 none &  2.99 \\
   $(\alpha(nn))(nn)$  &  2.97 \\
   $(((\alpha n)n)n)n$ &  unbound \\
   $((\alpha n)n)(nn)$ &  2.96 \\
   $((\alpha(nn))n)n$  &  2.96 \\
   $((tt)n)n$          &  2.36 \\
\end{tabular}
\end{table}
}

\noindent for the odd partial waves the realistic interaction of
Eikemeier and Hackenbroich \cite{Ei71} has been employed.
Both contain Coulomb, nuclear central, spin-orbit, and tensor 
components. The strength of the spin-orbit force was increased by a factor of
2.5.

\section{Momentum Distributions and Spatial Correlations}

In the simple picture of the serber model (sudden approximation) \cite{Se47} 
in which any influence of the
reaction mechanism of the fragmentation and the final-state interaction
is neglected, the measured momentum distribution is given by the 
Fourier transformation of the relative wave function between the fragments
integrated over the not observed momentum coordinates:
\begin{equation}
{{\rm d}N\over{\rm d}P_x} \propto
\int {\rm d}P_y {\rm d}P_z \; \rho_{\phi_1\phi_2}(\vec{P}),
\label{twoint}
\end{equation}
with
\begin{equation}
\rho_{\phi_1\phi_2}(\vec{P}) =
\left| {1\over(2\pi\hbar)^{3/2}} \int {\rm d}\vec{R}\; 
{\mit\Psi}_{\rm rel}(\vec{R})\;
\exp({{\rm i}\over \hbar}\vec{RP}) \right|^2 .
\end{equation}

The totally antisymmetrized relative wave function between the fragments
$|\phi_1\rangle$ and $|\phi_2\rangle$ is calculated as an overlap with the 
total wave function $|\phi_3\rangle$ where the integration is performed over all
coordinates except the relative coordinate $\vec{r}$
between the two fragments:
\begin{equation}
{\mit\Psi}_{\rm rel}(\vec{R}) = \langle \phi_1 \phi_2 |
    \delta (\vec{r}-\vec{R}) {\cal A} |\phi_3 \rangle.
\label{Psirel}
\end{equation}
In this sense the relative wave function is projected out of the 
multiconfiguration total wave function. This is necessary since the
different configurations are not orthogonal.
${\cal A}$ denotes the antisymmetrization operator.

The relative wave function can be projected onto a certain orbital angular
momentum $L$:
\begin{equation}
{\mit\Psi}^L_{\rm rel}(\vec{R})=Y_{LL}(\hat{\vec{R}})
\int {\rm d} \hat{\vec{R'}}\; Y^*_{LL}(\hat{\vec{R'}})\;
{\mit\Psi}_{\rm rel}(\vec{R'}).
\label{PsirelL}
\end{equation}

In the common reaction scenario one of the halo neutrons is removed by the 
target and scattered to large angles and therefore not detected. This
picture is based on the average neutron multiplicity measured to be close 
to one \cite{Ni96}. The remaining second neutron and the core may form
a short living resonant state. The measured momentum distribution of the
products of the decay of this resonance are thus no more only determined by
their original, internal motion in the halo nucleus. Rather the final-state
interaction has, at least in some cases, an important influence.   
The modification of the internal momentum distribution by the final-state 
interaction can be described 
\cite{Ko95,Zh94,Ni96} by using a
Breit-Wigner formula,
\begin{equation}
I_{\rm BW} = {1\over 2\pi}\;{{\mit\Gamma}\over (E-E_{\rm res})^2 +
	       {\mit\Gamma}^2/4} \;,
\label{BreitWigner}
\end{equation}
\vbox{
\begin{figure}
\centerline{\epsfxsize=8.5cm  \epsfbox{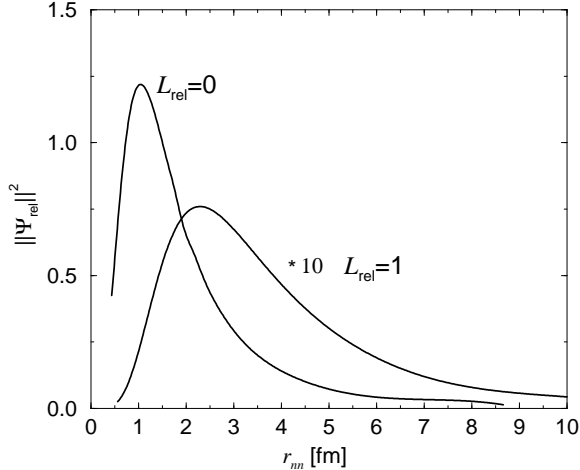}}
\caption{Square of the norm of the projected relative wave function 
	  $\|{\mit\Psi}^L_{\rm rel}\|^2$ as a function of the r.m.s. 
	  radius of 
	  $|nn\rangle$.}
\label{he6nncor}
\end{figure}
}
with resonance energy $E_{\rm res}$ and width $\mit\Gamma$. 

For the calculation of the momentum distribution of the $^4$He fragments 
from $^6$He breakup one
has $|\phi_1\rangle=|^4{\rm He}\rangle$, $|\phi_2\rangle=|nn\rangle$ and
$|\phi_3\rangle=|^6{\rm He}\rangle$. The $|^4{\rm He}\rangle$ function 
was determined by variation and has, in good agreement with experimental data,
a point nucleon root mean square radius of 1.42 fm. The $\alpha$ cluster
radii inside $|^6{\rm He}\rangle$ are very similar to this value but the 
momentum
distribution is not very sensitive to it.

To determine
the function $|nn\rangle$ and thus the correlation of the two halo neutrons
in $^6$He we took the norm of the projected
relative wave function $\|{\mit\Psi}_{\rm rel}\|$ as
a measure of the content of a certain $|nn\rangle$ configuration
in $|^6{\rm He}\rangle$ for a fixed $|^4{\rm He}\rangle$ fragment.
Figure \ref{he6nncor} shows $\|{\mit\Psi}^L_{\rm rel}\|^2$ as a function of the 
root mean 
square radius of $|nn\rangle$ which is a good parametrization of the
simple Gaussian function which has been used for the $\vec{r}$-space 
representation of $|nn\rangle$. 
The two pronounced peaks in Fig.\ \ref{he6nncor}, at 1.0 fm for $L_{\rm rel}=0$ and
at 2.3 fm for $L_{\rm rel}=1$, correspond to a spatialy strongly correlated 
dineutron configuration and a spatialy strongly anticorrelated cigar-like
neutron pair. This observation is in agreement with the results
of a hyperspherical harmonics calculation (HH) \cite{Zh91}.

The $|nn\rangle$ configurations with maximal $\|{\mit\Psi}^L_{\rm rel}\|$
have been choosen for the calculation of ${\mit\Psi}_{\rm rel}(\vec{R})$. 
The resulting momentum distribution for $^4$He fragments in sudden
approximation is compared in Fig.\
\ref{he6mom} with the
measured transverse momentum distribution \cite{Ko92}
and a calculation in a HH 
model \cite{Zh91} where also the approximation of the serber model was used.
Both theoretical curves are in good agreement with the experiment. 
This indicates that, besides the fulfilled
preconditions for the serber model, high energy, light target and a
loosely bound projectile, the $^4$He\thinspace$n$ final-state interaction
hardly effects the $^4$He momentum distribution. Also the $^4$He target
interaction seems to play no important role for the detected $^4$He 
\vbox{
\begin{figure}
\centerline{\epsfxsize=8.5cm  \epsfbox{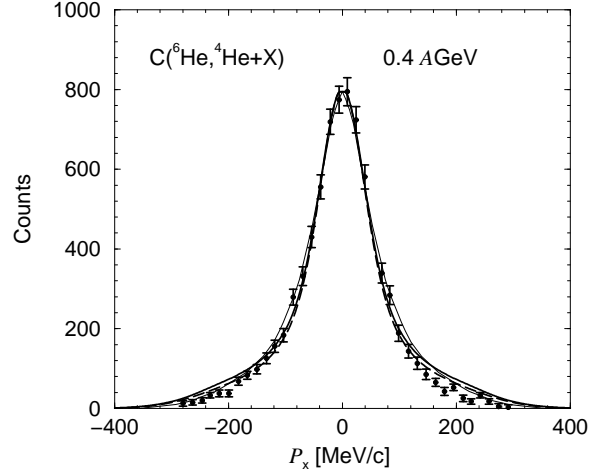}}
\caption{$^4$He transverse momentum distribution from a $^6$He breakup 
         calculated in the RRGM with final state interactions (thick solid 
	 line) and in sudden approximation (dashed line) 
	 in comparison with experimental data \protect\cite{Ko92}
	 and the HH model \protect\cite{Zh91} in sudden 
	 approximation (thin solid line).}
\label{he6mom}
\end{figure}
}
fragments.

Taking final-state interactions
into account, 
the $^4$He transverse momentum distribution
is given by
\begin{eqnarray}
{{\rm d}N^{\rm FSI}_{^4{\rm He}} \over {\rm d}P_x} &\propto&
\int {\rm d}P_y{\rm d}P_z \int {\rm d}\vec{P}'{\rm d}\vec{P}'' \;
\rho_{^4{\rm He}(nn)}(\vec{P}) \nonumber \\
 & & I_{\rm BW}(\vec{P}') \;
\delta(\vec{P}'+{4\over5}\vec{P}''+\vec{P}),
\label{he4momdis}
\end{eqnarray}
where $\vec{P}''$ is the momentum of the removed neutron and
$\vec{P}$ the $^4$He momentum, both in the $^6$He c.m.\ system
(cf.\ Fig.\ \ref{Mom}).
$\vec{P}'$ is the relative momentum of the detected neutron and the $^4$He
fragment in the $^5$He c.m.\ system, connected with $E$ from Eq.\ 
(\ref{BreitWigner}) by
\begin{equation}
E={P'^2\over2{5\over4}m} \;,
\end{equation}
with the nuclear mass $m$. We used $E_{\rm res}=0.89$ MeV and 
${\mit\Gamma} = 0.6 $ MeV in Eq.\ (\ref{BreitWigner})
for the $^5$He ground state resonance \cite{Aj88}.
Indeed Fig.\ \ref{he6mom} shows that the final-state 
\vbox{
\begin{figure}
\centerline{\epsfxsize=5.7cm  \epsfbox{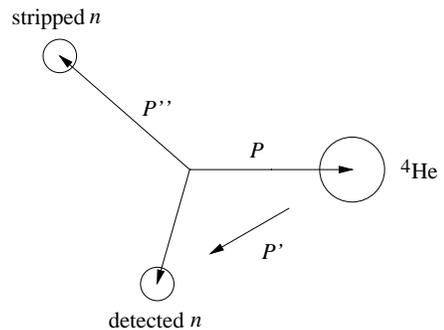}}
\caption{Definition of the momenta appearing in Eq.\ (\protect\ref{he4momdis}).}
\label{Mom}
\end{figure}
}
\vbox{
\begin{figure}
\centerline{\epsfxsize=8.5cm  \epsfbox{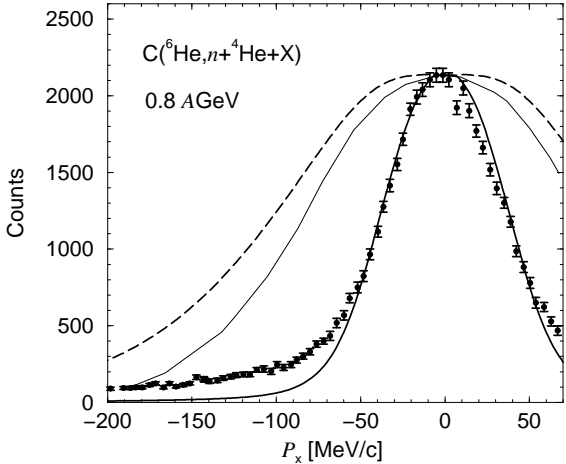}}
\caption{Neutron transverse momentum distribution from a $^6$He breakup 
         calculated in the RRGM with final state interactions (thick solid 
	 line) and in sudden approximation (dashed line) 
	 in comparison with experimental data \protect\cite{Ko93}
	 and the HH model \protect\cite{Ko94} in sudden 
	 approximation (thin solid line).}
\label{n_mom}
\end{figure}
}
interaction does 
almost not effect the momentum distribution of the $^4$He fragments.

In order to calculate the neutron momentum distribution one has to employ
$|\phi_1\rangle = |^5{\rm He}\rangle$ and $|\phi_2\rangle = |n\rangle$.
The wave function for the unbound system $|^5{\rm He}\rangle$ was constructed
in the same way as for the dineutron $|nn\rangle$. The maximum of
$\|{\mit\Psi}_{\rm rel}\|$ was determined with respect to the relative
wave function between the $\alpha$ cluster and the neutron in 
$|^5{\rm He}\rangle$ for a
fixed $\alpha$. In this sense the mean value of the $\alpha$ core
halo neutron
radius inside $^6$He was determined to 2.0 fm. The corresponding
momentum distribution (cf.\ Fig.\ \ref{n_mom}) in sudden approximation 
is much broader than the measured 
transverse neutron momentum distribution \cite{Ko93} but in agreement with
a three-body calculation in the serber model \cite{Ko94}.

Some different processes might be responsible for this discrepancy.
The $^4$He\thinspace$n$ final-state interaction is considered to be the
most dominant one \cite{Ko94}. But moreover, 
in contrast to the $^4$He fragments the detected neutrons
might have different origins: they could be emitted from the neutron halo
by a direct breakup or with a target interaction, they could be emitted
from the $^4$He core or even from the target itself \cite{Ko94}.

Taking final-state interactions
into account, 
the neutron transverse momentum distribution is given by
\begin{eqnarray}
{{\rm d}N^{\rm FSI}_{n} \over {\rm d}P_x} &\propto&
\int {\rm d}P_y{\rm d}P_z \int {\rm d}\vec{P}'{\rm d}\vec{P}'' \;
\rho_{^5{\rm He}\,n}(\vec{P}'') \nonumber \\
 & & I_{\rm BW}(\vec{P}') \;
\delta(\vec{P}'-\vec{P}-{1\over5}\vec{P}''),
\label{he6_n_FSI}
\end{eqnarray}
where $\vec{P}$ is the momentum of the detected neutron.
The meaning of $\vec{P}'$ and $\vec{P}''$ is the same as in Eq.\
\ref{he4momdis} (cf.\ Fig.\ \ref{Mom}).
Contrary to the $^4$He fragment, the neutron momentum 
\vbox{
\begin{figure}
\centerline{\epsfxsize=8.5cm \epsfbox{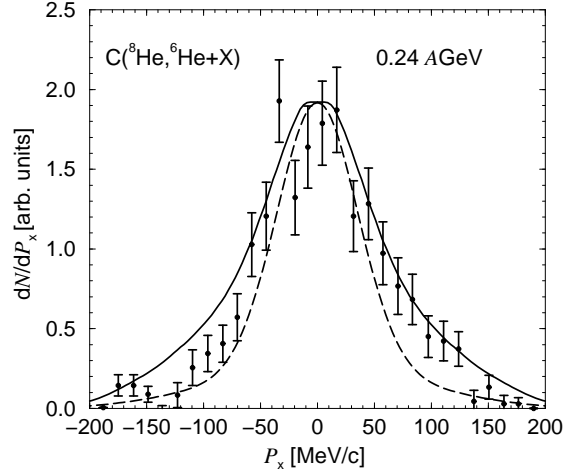}}
\caption{$^6$He transverse momentum distribution from $^8$He breakup 
	 calculated in the RRGM with final-state interactions (solid line)
	 and in sudden approximation (dashed line)
	 in comparison with experimental data \protect\cite{Ni96}.}
\label{he8mom}
\end{figure}
}
distribution is
considerably influenced by the final-state interaction 
(cf.\ Fig.\ \ref{n_mom}). The distribution (\ref{he6_n_FSI}) is in
good agreement with the measured distribution. The remaining discrepancy
beyond 60 ${\rm MeV}\over c$ might be due to neglecting the
neutron target interaction \cite{Ko94}.

The momentum distribution for $^6$He fragments from $^8$He breakup was    
calculated in a way analogical to the determination of the $^4$He distribution
from $^6$He breakup:
The determination of maxima of $\|{\mit\Psi}^L_{\rm rel}\|$ showed
that the second halo neutron pair has a strong correlated component
with a root mean square radius of 1.2 fm for spin $s=0$ and a less 
correlated component with a radius of 1.8 fm for spin $s=1$.
The momentum distribution for $^6$He fragments 
calculated
in the RRGM model in sudden approximation seems to be a bit to narrow but 
is still compatible with the measured 
transverse momentum distribution \cite{Ni96} 
(cf.\ Fig.\ \ref{he8mom}). 

Taking final-state interactions
into account,
the $^6$He transverse momentum distribution
is given by
\begin{eqnarray}
{{\rm d}N^{\rm FSI}_{^6{\rm He}} \over {\rm d}P_x} &\propto&
\int {\rm d}P_y{\rm d}P_z \int {\rm d}\vec{P}'{\rm d}\vec{P}'' \;
\rho_{^6{\rm He}(nn)}(\vec{P}) \nonumber \\
 &  & I_{\rm BW}(\vec{P}') \;
\delta(\vec{P}'+{6\over7}\vec{P}''+\vec{P}),
\label{he6momdis}
\end{eqnarray}
where the momenta $\vec{P}$, $\vec{P}'$ and $\vec{P}''$ are defined 
analogically to Eq.\ (\ref{he4momdis}).

For the parameters $E_{\rm  res}$ and $\mit\Gamma$ in Eq.\ (\ref{BreitWigner})
we employed not only the experimental values for the $^7$He ground state
resonance (0.44 MeV and 0.16 MeV) \cite{Aj88} but also the results of an 
S-matrix
analysis of the phase shifts calculated in the RRGM (0.442 MeV and
0.114 MeV) in \cite{Wu97}. The resulting momentum distributions differ
only within the line width in Fig.\ \ref{he8mom} and \ref{he8_n_mom}.
It is shown in Fig.\ \ref{he8mom} that the final-state 
interaction increases the width of the $^6$He 
\vbox{
\begin{figure}
\centerline{\epsfxsize=8.5cm  \epsfbox{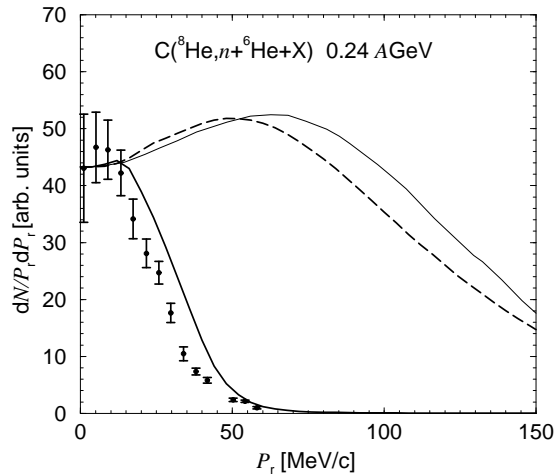}}
\caption{Neutron radial momentum distribution from $^8$He breakup
	 calculated in the RRGM with final-state interactions (thick solid 
	 line) and in sudden approximation (dashed line)
	 in comparison with experimental data \protect\cite{Ni96}
	 and COSMA calculation 
	 \protect\cite{Ni96} in sudden approximation (thin solid line).}
\label{he8_n_mom}
\end{figure}
}
momentum distribution in such a way
as to yield good agreement with the measured one.
 
The neutron momentum distribution was calculated by setting $|\phi_1\rangle=
|^7{\rm He}\rangle$ and $|\phi_2\rangle=|n\rangle$. The maximum of
$\|{\mit\Psi}_{\rm rel}\|$ was determined with respect to the relative
wave function between the $^6$He function and the neutron in 
$|^7{\rm He}\rangle$ for a
fixed $^6$He. The mean value of the $^6$He\thinspace$n$ radius was determined
to 2.3 fm. Contrary to Eq.\ (\ref{twoint}) the momentum 
distribution for neutrons
from $^8$He breakup has been calculated by integration only over one
momentum coordinate in order to compare with experimental data
\cite{Ni96}. The RRGM and the COSMA calculations \cite{Ni96} in sudden 
approximation
are in good agreement (cf.\ Fig.\ \ref{he8_n_mom}) with each other but 
much broader than
the measured radial momentum distribution. This discrepancy is
presumable due to the same reasons as for the $^6$He breakup.

Taking final-state interactions
into account,
the neutron radial momentum distribution is given by
\begin{eqnarray}
{{\rm d}N^{\rm FSI}_{n} \over P_{\rm r}\,{\rm d}P_{\rm r}} &\propto&
\int {\rm d}P_z \int {\rm d}\vec{P}'{\rm d}\vec{P}'' \;
\rho_{^7{\rm He}\,n}(\vec{P}'') \nonumber \\
 & & I_{\rm BW}(\vec{P}') \;
\delta(\vec{P}'-\vec{P}-{1\over7}\vec{P}'').
\label{he6_n_mom_FSI}
\end{eqnarray}
The momenta $\vec{P}$, $\vec{P}'$ and $\vec{P}''$ are defined 
analogically to Eq.\ (\ref{he6_n_FSI}).
The extracted width of 66 ${\rm MeV}\over c$ is a little larger than the
measured one of 54 ${\rm MeV}\over c$ (cf.\ Fig.\ \ref{he8_n_mom}).

The origin of this small deviation from the experimental data is not obvious.
The width of the distribution given by Eq.\ (\ref{he6_n_mom_FSI}), 
however, depends very strongly on the position of
the $^7$He resonance but not on the resonance width. For example lowering
the parameter $E_{\rm res}$ in Eq. (\ref{BreitWigner}) by about 100 keV
yields perfect agreement with the data (see also \cite{Ni96}).

\section{Summary}

It was shown that the ground state properties of halo nuclei and the
momentum distributions of fragments and neutrons from breakup reactions can 
be well reproduced in a consistent way in the microscopic description
of the refined resonating group method.
The momentum distributions of $^4$He and $^6$He fragments from
$^6$He and $^8$He breakup, respectively,  
are in good agreement with the measured 
distributions and almost not influenced by final-state interactions.
On the contrary, the neutron momentum distributions can only be explained
by taking final-state interactions into account.

\acknowledgements{This work was supported by DFG and BMBF.}

\end{document}